\documentclass[twocolumn]{aastex63}
\usepackage{textcomp}
\usepackage{bm}
\usepackage{enumitem}
\usepackage[table,xcdraw]{colortbl}
\setitemize{noitemsep,topsep=0pt,parsep=0pt,partopsep=0pt}

\begin{document}
\title{Galactic Mass Estimates using Dwarf Galaxies as Kinematic Tracers}
\author{Anika Slizewski}
\email{aslizew@uw.edu}
\affiliation{Department of Astronomy, University of Washington, Seattle, WA}
\author{Xander Dufresne}
\affiliation{University of Toronto, Toronto, Canada}
\author{Keslen Murdock}
\affiliation{University of Toronto, Toronto, Canada}
\author[0000-0003-3734-8177]{Gwendolyn Eadie}
\email{gwen.eadie@utoronto.ca}
\affiliation{David A.~Dunlap Department of Astronomy \& Astrophysics, University of Toronto, Toronto, Canada}
\affiliation{Department of Statistical Sciences, University of Toronto, Toronto, Canada}
\author[0000-0003-3939-3297]{Robyn Sanderson}
\affiliation{University of Pennsylvania, Philadelphia, PA}
\author[0000-0003-0603-8942]{Andrew Wetzel}
\affiliation{Department of Physics \& Astronomy, University of California, Davis, CA}
\author{Mario Juric}
\affiliation{Department of Astronomy, University of Washington, Seattle, WA}

\submitjournal{ApJ}

\received{July 22, 2021}

\begin{abstract}
    New mass estimates and cumulative mass profiles with Bayesian credible regions (c.r.) for the Milky Way (MW) are found using the Galactic Mass Estimator (GME) code and dwarf galaxy (DG) kinematic data from multiple sources. GME takes a hierarchical Bayesian approach to simultaneously estimate the true positions and velocities of the DGs, their velocity anisotropy, and the model parameters for the Galaxy's total gravitational potential. In this study, we incorporate meaningful prior information from past studies and simulations. The prior distributions for the physical model are informed by the results of \citet{Eadie_2019}, which used globular clusters instead of DGs, as well as by the subhalo distributions of the Ananke Gaia-like surveys from Feedback In Realistic Environments-2 (Fire-2) cosmological simulations \citep[see][]{Sanderson_20}. Using DGs beyond 45 kpc, we report median and 95\% c.r estimates for $r_{200}$ = 212.8 (191.12,238.44) kpc, and for the total enclosed mass $M_{200}$ = 1.19 (0.87,1.68)$\times10^{12}M_{\odot}$ (adopting $\Delta_c=200$). Median mass estimates at specific radii are also reported (e.g., $M(<50\text{ kpc})=0.52\times10^{12}M_{\odot}$ and $M(100\text{ kpc})=0.78\times10^{12}M_{\odot}$). 
    Estimates are comparable to other recent studies using {\it Gaia} DR2 and DGs, but notably different from the estimates of \citet[]{Eadie_2019}. 
    We perform a sensitivity analysis to investigate whether individual DGs and/or a more massive Large Magellanic Cloud (LMC) on the order of $10^{11}M_{\odot}$ may be affecting our mass estimates. We find possible supporting evidence for the idea that some DGs are affected by a massive LMC and are not in equilibrium with the MW.
\end{abstract}


\section{Introduction}\label{sec:intro}
The total mass of a galaxy is an important quantity for understanding astronomical theories about dark matter halos and galactic evolution. Photometric observations are commonly used to constrain a galaxy's luminous mass, but the development of mass estimators based on kinematic data leads to a more comprehensive analysis of a galaxy and its dark matter halo. In the case of our own Milky Way (MW) Galaxy, complete data on the motions of dwarf galaxies (DGs) and other kinematic tracers are available. These satellites are located in the MW's halo and can be used to constrain the total mass distribution. 

In a series of papers \citep[][hereafter Papers I-V]{Eadie_15,Eadie_16,Eadie_17,Eadie_18,Eadie_2019}, a hierarchical Bayesian method was developed for estimating cumulative mass profiles (CMPs) of the MW using kinematic data from tracers. Paper V most recently applied this method to globular cluster (GC) data from \citet[Gaia DR2]{GAIA_2016,Gaia_18} and the catalog by \citet{Vasiliev_19,Vasiliev_19_catalog} and found $M_{200}=0.70\times 10^{12} M_{\odot}$, corresponding to the mass at the radius $r_{200}$. This result is on the lower end of recent estimates for the Galaxy's total mass (e.g. \citet{Monari_18,Vasiliev_19,Posti_2019}). At the same time, it is notable that Paper V found $M(50\text{ kpc})=0.37(0.29,0.51)\times 10^{12} M_{\odot}$ (50\% credible regions in brackets), in agreement with many other studies reporting mass estimates at that distance \citep[e.g.]{Kochanek_96,Deason_12,Gibbons_14,Erkal_19}. Thus, there is still a disagreement in the mass estimates within larger distances of the MW and the total mass (which is often reported as $M_{200}$, with $\Delta_c=200$ and $\rho(r<r_{vir})=\Delta_c\rho_c$).

Our lower mass estimates in previous papers could be attributed to a few factors discussed in papers III-V and reiterated briefly here. First, the choice of a power-law model for the gravitational potential does not account for the bulge and disk of the MW. Second, the choice of tracer could play a role in the mass estimate; the majority of GCs are located within 50kpc of the Galactic center (Figure~\ref{fig:datargcs}), making it difficult to constrain the mass out to larger distances. Third, the Bayesian approach is possibly biased to lower mass estimates; a slight systematic bias in the median estimate of $M_{200}$ was observed when testing the procedure on mock data from 18 galaxies generated by the McMaster Unbiased Galaxy Simulations 2 (see Paper IV). However, the sample size of galaxies in that study was quite small, and the 95\% credible regions  on $M_{200}$ still contained the true value for almost all cases. Paper IV also showed that the inner and outer regions of the mock galaxies' CMPs could not be accurately estimated \textit{simultaneously} with the simulated GC data, which may have led to the observed small bias. While a more thorough study of a larger sample of mock galaxies might help us fully understand any potential bias in $M_{200}$ with our Hierarchical Bayesian method, such a study would rest heavily upon the specific simulations used.

Instead of relying solely on simulations, many studies now combine the power of simulations with real data. For example, the galactic potential and tracer distribution models used in many of the studies reviewed in \citet{Wang_20} are calibrated against N-body and hydrodynamical cosmological simulations (e.g. Bolshoi, Illustris, EAGLE, Millennium), calling for simulations that accurately represent MW-like galaxies. Simulations incorporate our current understanding of the physics within galaxies and their dark matter halos, and therefore can be used to set informative priors in a Bayesian analysis of real data.

In this work, we revisit the Galaxy mass estimation problem with the hierarchical Bayesian method of Paper V. We use the compiled data set of MW dwarf galaxies (DGs) presented in \citet{Riley_19}, which includes proper motions measured by Gaia DR2 proper motions \citep{Gaia_18}, and we inform our prior distributions using both simulations of MW-like galaxies and the marginal posterior distributions from Paper V. \textit{By including both lines of prior information, we hope to achieve a good balance between our current understanding of the MW's dark matter halo via simulations and observations.} MW satellites provide information out to larger distances than GCs. Moreover, at these larger distances, our model assumption of a power-law potential for the Galaxy's outer halo may be seen as less egregious. We benefit and draw from the previous results of Paper V, using the marginal posterior distributions for the gravitational potential model parameters ($\Phi_0$ and $\gamma$, see Section~\ref{sec:massmodel}).

One of our main motivations for revisiting the MW mass estimation problem is that recent studies have suggested the Large Magellanic Cloud (LMC) may be on its first pass of the MW and is much more massive than previously thought  \citep[possibly on the order of $1\textnormal{-}2.5\times10^{11}M_{\odot}$, e.g.,][]{Besla_15,Laporte_2018}. 

As the LMC infalls, it may be influencing the orbits of some MW DGs, thus challenging the assumption of tracer independence and boundedness in our model and others. For example, \citet{Erkal_20} used equilibrium models with recent mass estimates for the LMC and found a significant bias on the MW mass estimate, showing overestimation of the MW mass when ignoring or using higher masses for the LMC. Additionally, \citet{Boubert_20} investigated how the LMC has deflected the orbits of hypervelocity stars, which are a very common type of tracer to use as a probe for the MW's mass and shape. Investigating which tracers may affect Galaxy mass estimates the most could provide evidence (or not) for a more massive LMC.

Papers I-V rely on the assumption that the tracers are not only independent but also bound to the Galaxy. This assumption is likely valid for GCs, but may not be true for all DGs around the MW. Should any DGs be unbound, we might expect an inflated mass estimate under our model assumptions. For this reason, we not only perform an analysis of the MW's cumulative mass profile using the entire DG sample, but also investigate how individual DGs affect our mass estimates through a sensitivity analysis.

Thus, this paper seeks to answer the following questions:
\begin{enumerate}
    \item \textit{What is the MW cumulative mass profile when using DG data?}
    \item \textit{Does our model still produce lower estimates when using DG data?}
    \item \textit{How sensitive are the mass estimates to particular DGs that could be affected by a massive LMC? }
\end{enumerate}

Other studies have already make great use of the second \textit{Gaia} data release \citep[Gaia DR2, see][and references therein]{GAIA_2016,Gaia_18,Gaia_18D} and the subsequent updated position and velocity data of DGs around the MW. Before jumping into our analysis, we briefly list some of the most recent literature on MW mass estimates. For example,

\begin{itemize}
    \item \citet[]{Cautun_20} found 68\% credible regions (c.r.) $M_{200} = 1.08^{+0.20}_{-0.14}\times 10^{12}M_{\odot}$. They compared estimates using a contracted halo and a generalized Navarro-Frenk-White (NFW) halo profile, and analyzed baryonic and DM mass in simulations based on Gaia DR2 data. A discrepancy between the different gravitational models was only observed within about 30 kpc. This provides supporting evidence that a simple NFW profile is a valid representation of the MW mass distribution beyond 30kpc. 

    \item \citet{Watkins_2019} used the tracer-mass-estimator developed in \citet{Watkins_2010} on Gaia DR2's GC kinematics and found an estimate of $M(<21.1\text{ kpc})=0.21^{+0.03}_{-0.04}$. This is quite similar to Paper V's results of 50\% c.r. $M(<21.1\text{ kpc})=0.24(0.2,0.27)$. 

    \item \citet{Callingham_19} applied the simulation-based distribution function method developed in \citet{Li_2017} with the EAGLE and Auriga simulations and Gaia DR2 classical satellite data to estimate 68\% c.r. $M_{200}=1.17^{+0.21}_{-0.15}\times 10^{12}M_{\odot}$. 

    \item \citet[]{Li_2020} recently updated their method and found a similar estimate of $M_{200}=1.23^{+0.21}_{-0.18}\times 10^{12}M_{\odot}$, and $M_{200}=1.26^{+0.17}_{-0.15}\times 10^{12}M_{\odot}$ when including halo stars.

    \item \citet{Deason_19} used Gaia DR2's high velocity halo stars and found $M_{200} = 1.00^{+0.31}_{-0.24}\times 10^{12}M_{\odot}$. One of their model priors on the velocity distribution had strong influence and increased their estimate -- the biases were investigated in \cite{Grand_2019}, and the result increased to $M_{200} = 1.29^{+0.37}_{-0.47}\times 10^{12}M_{\odot}$. 
    
    \item \cite{Monari_18} also used the kinematics of Gaia DR2's halo stars with a velocity power-law model to obtain $M_{200} = 1.28^{+0.68}_{-0.50}\times 10^{12}M_{\odot}$. In related work, \cite{Hattori_18} investigated the metal-poor, extreme velocity stars found in Gaia DR2 and produced a rough estimate of $M_{200} \approx1.4\times 10^{12}M_{\odot}$. 

    \item \citet{Riley_19} also compiled a catalog of MW DG satellites with all of the necessary information to estimate the MW mass within the hierarchical Bayesian framework of Papers I-V. They investigated the velocity anisotropy of the inner and outer regions of the MW using Gaia DR2 tracer data and the APOSTLE and Auriga simulations. 

\end{itemize}

We will revisit these results in light and in comparison to our study in the discussion portion of this paper. The paper is structured as follows: Section~\ref{sec:data} describes the tracers used in the analysis. Section~\ref{sec:methods} reviews the analysis methods and the hierarchical Bayesian model. Section~\ref{sec:physmodel} describes how the CMP is calculated, and Section~\ref{sec:priors} describes how we define the prior and hyperprior distributions. Section~\ref{sec:results} presents the Bayesian estimates, the CMP, and $M_{200}$ for the entire set of data used in our study, while Section~\ref{sec:subsets} presents a sensitivity analysis on different subsets of the data. A discussion is also included throughout these two sections. We conclude and present ideas for future work in Section~\ref{sec:conc}.

\section{Data}\label{sec:data}

\begin{figure}
    \centering
    \includegraphics[width=\columnwidth]{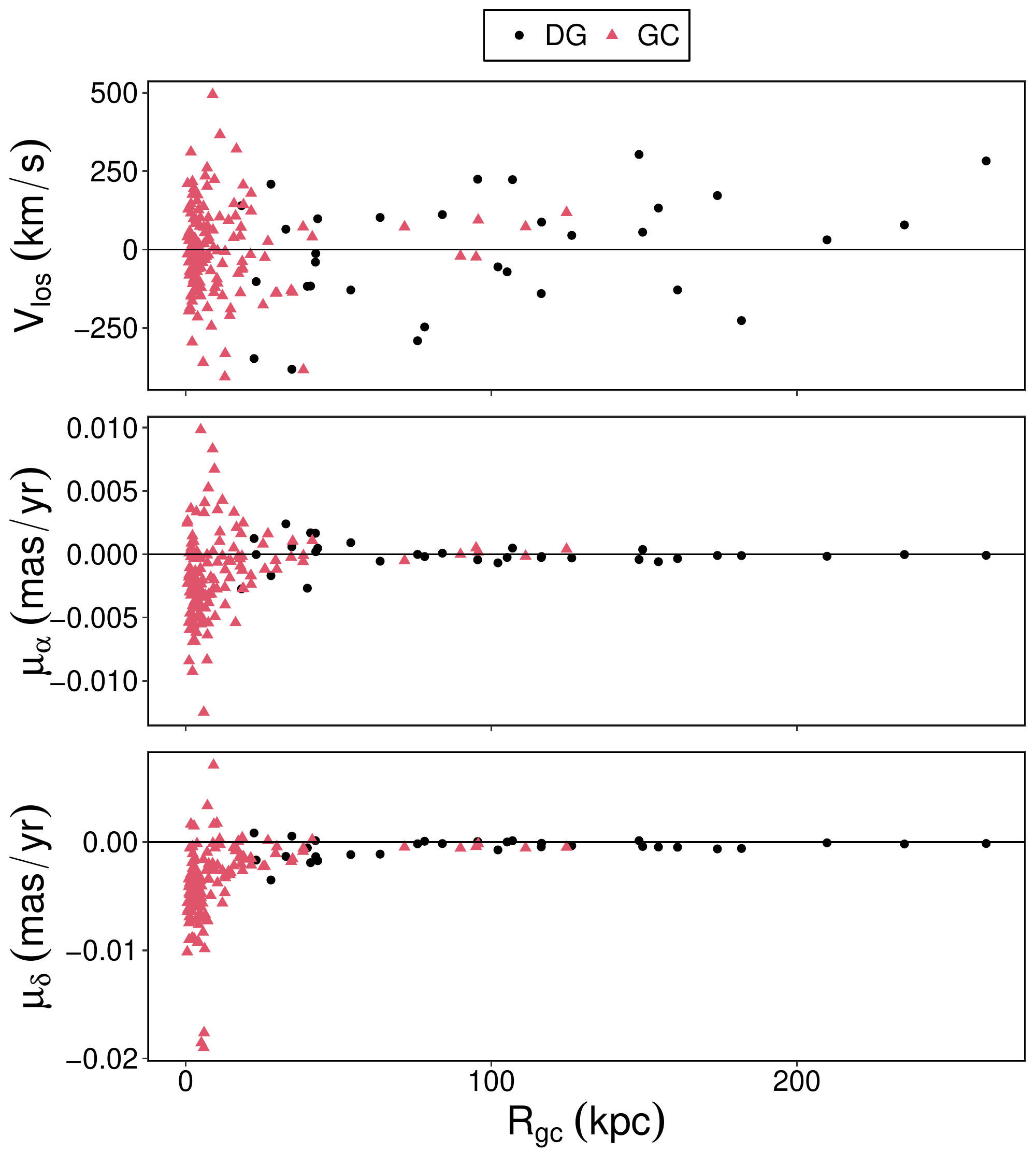}
    \caption{Line-of-sight velocities and proper motions for the DGs data set (black circles) compared to the GCs used in Paper V (red triangles). The DGs extend much farther and also have a smaller range of proper motions.}
    \label{fig:datargcs}
\end{figure}

In this work, we use the list of Galactic dwarf satellites compiled by \citet{Riley_19}\footnote{\url{https://github.com/ahriley/beta-MW-dwarfs}} (see Figure~\ref{fig:datargcs}). These data consist of 36 tracers, including ten measured by \textit{Gaia} DR2 \citep{Gaia_18,Gaia_18D}, and a few others that are close to or associated with the Magellanic Clouds. The tracers are located beyond 15 kpc and extend to about 260 kpc. Figure!\ref{fig:datargcs} shows the spatial extent of the DG sample (black circles) compared to the extent of the GCs (red triangles). 

The summary of \textit{Gaia} DR2 data and instrumentation is detailed in \cite{GAIA_2016,Gaia_18}. We exclude the SMC and four LMC-associated DGs  (Carina II, Carina III, Horologium I and Hydrus I) outlined in \cite{Kallivayalil_18} and \citet{Patel_20} from our analysis.

\section{Methods}\label{sec:methods}
We employ the Galactic Mass Estimator (GME) code to estimate the mass of the Milky Way (MW) using the kinematic information of dwarf galaxies as tracers. The hierarchical Bayesian model of GME was developed in Papers I-V, and here we use the previous results from Paper V to define hyperprior distributions for two of the model parameters. In Section~\ref{sec:subsets} we also perform sensitivity analysis on the results to investigate bias in the methods. 

The distribution function for Galactic tracers used in GME was derived in \citet{Evans_97}, and the CMP is given in \citet{Deason_12} and \citet{Watkins_2010}. Measurement uncertainties were introduced via the hierarchical model in Paper II and III, and is included henceforth. These papers focused on the use of GCs from the \cite{harris2010} catalog. The measurements of Galactocentric position $r$ (kpc) and heliocentric velocities $\mu_{\alpha}$ (arcsec/yr), $\mu_{\delta}$ (arcsec/yr), and $v_{los}$(km/s) are assumed to have Gaussian distributions centered at their true values, with standard deviations equal to the measurement uncertainties. Details about the measurement model, including the conventions of solar motion and conversions to Galactocentric coordinates, are in Papers I-IV, especially in Paper III.

GME is written in the R Statistical Software Language and is openly available\footnote{\url{https://github.com/gweneadie/GME}, although a faster version using Stan will be available in a forthcoming paper (Shen et al, in prep)}.

\subsection{Physical Model}\label{sec:physmodel}
The model in GME assumes a total gravitational potential 
\begin{equation}
\Phi(r) = \Phi_{0}r^{-\gamma},
\label{eqn:gp}
\end{equation}
where $\Phi_0$ and $\gamma$ are parameters, and $r$ is the Galactocentric distance. The density profile of the tracer population $\rho(r)$ is assumed to follow
\begin{equation}
    \rho \propto r^{-\alpha},
    \label{eqn:sd}
\end{equation}
where $\alpha$ is a parameter. The tracer population is also assumed to have a constant anisotropy $\beta$. Thus, the model parameters that are:
\begin{equation}
    \bm{\theta} = (\Phi_{0},\gamma,\alpha,\beta)
\label{eqn:theta}
\end{equation}
These are, respectively, the scale and power law slope of the gravitational potential, the power law slope of the tracer distribution, and the velocity anisotropy parameter.

The kinematic tracers (in this paper, DGs) are assumed to be independent and bound to the Galaxy. The true but unknown Galactocentric position $r$ and velocity $v$ parameters, along with the model parameters, determine the specific energy $\mathcal{E}$ and angular momentum $L$ of each tracer:

\begin{equation}\label{eqn: energy}
    \mathcal{E} = -\frac{1}{2}({v_r}^{2}+{v_t^2})+\Phi(r)
\end{equation}
and 
\begin{equation}
    L = rv_{t}
\end{equation}

Following equations~\ref{eqn:gp} and \ref{eqn:sd}, the distribution function for the specific energy and angular momentum $f(\mathcal{E,L})$ of a tracer is given by
\begin{equation}
    f(\mathcal{E}, L) = \eta L^{-2\beta}\mathcal{E}^{\frac{\beta(\gamma-2}{\gamma}+\frac{\alpha}{\gamma} - \frac{2}{3}}
\end{equation}
where
\begin{equation}
    \eta = \frac{ \Phi_0^{\frac{2\beta}{\gamma}-\frac{\alpha}{\gamma}} \Gamma\left( \frac{\alpha}{\gamma} - \frac{2\beta}{\gamma} + 1 \right)}{\sqrt{8\pi^3 2^{-2\beta}} 
    \Gamma\left(1-\beta\right) \Gamma\left(\frac{\beta(\gamma-2)}{\gamma} + \frac{\alpha}{\gamma} - \frac{1}{2}\right)}
\end{equation}
is the normalizing constant found from equation 3.7 in \cite{Evans_97}.

The true Galactocentric positions and true heliocentric velocities are treated as parameters in the hierarchical model; the measurements are assumed to be drawn from normal distributions with mean equal to the true value and the standard deviation equal to the measurement uncertainty. In this way, the true positions and velocities are also estimated via Bayesian inference (see Papers III-IV for more details). 

\subsection{Prior and Hyperprior Distributions}\label{sec:priors}
Bayesian inference requires priors on the model parameters, which are summarized in Table~\ref{table:priors} and discussed below.

\begin{figure}[ht!]
    \begin{center}
    \includegraphics[width=\columnwidth]{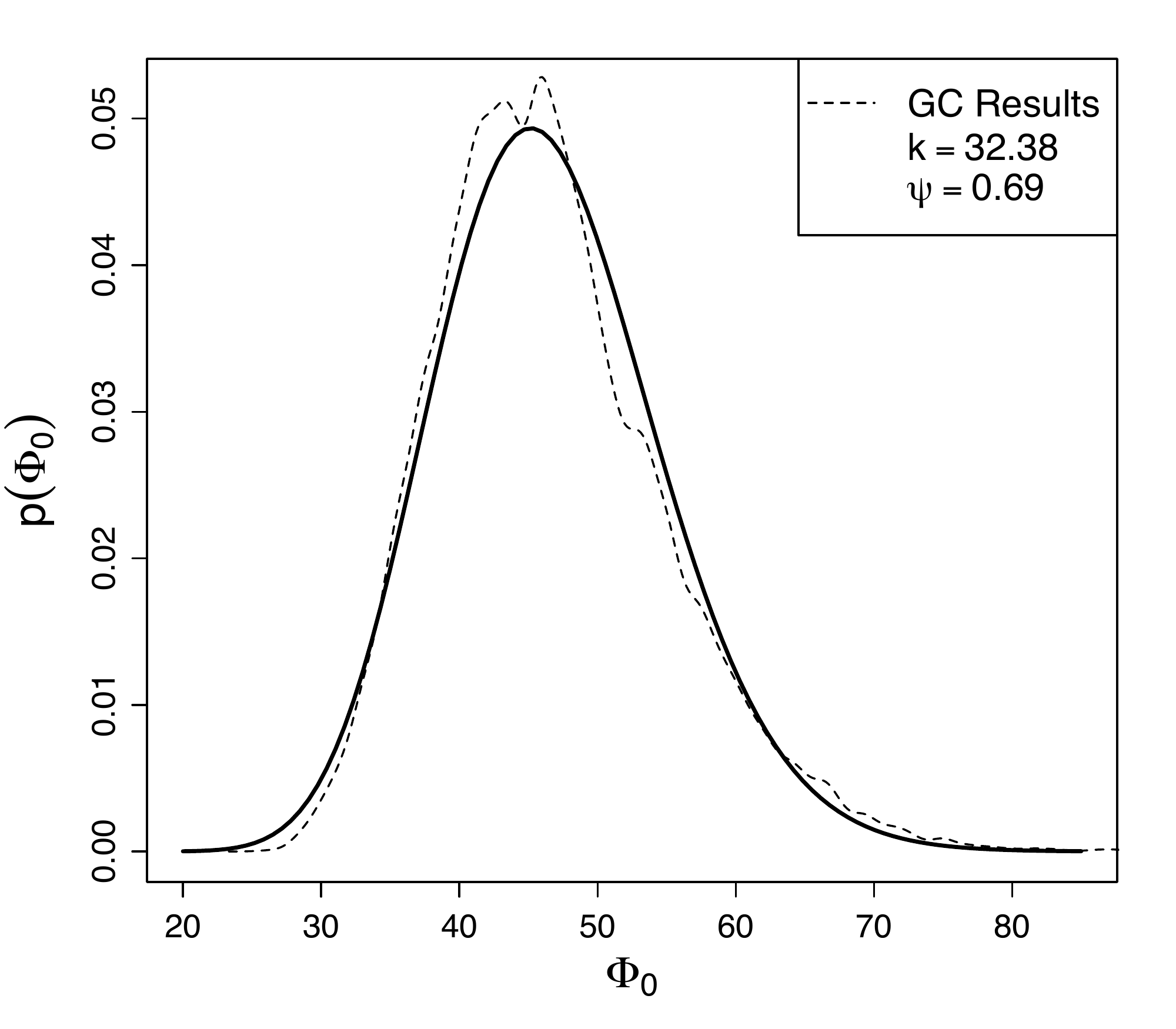}
    \caption{Prior distribution for the gravitational potential parameter $\Phi_0$ (solid). Gamma distribution fit to the marginal distribution of $\Phi_0$ from Paper V (dashed), with shape $k=32.38$ and rate $\psi=0.69$.}
    \label{fig:phi0}
    \end{center} 
\end{figure}

\begin{figure}[ht!]
    \begin{center}
    \includegraphics[width=\columnwidth]{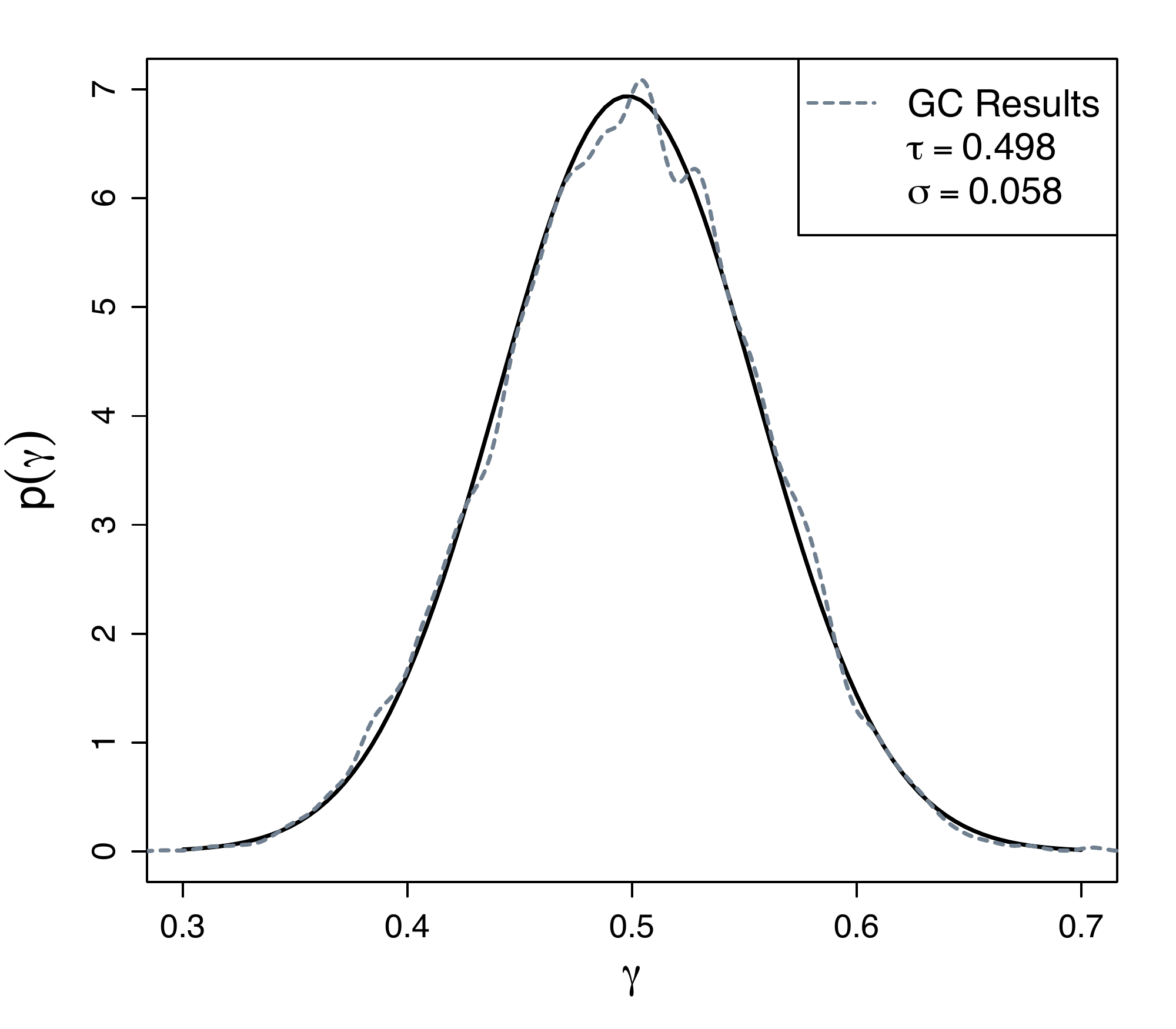}
    \caption{Prior distribution for the potential power law slope $\gamma$ (solid curve). The Gaussian distribution was fit to the marginal distribution of $\gamma$ from Paper V (dashed), with mean $\tau=0.498$ and standard deviation $\sigma=0.058$.}
    \label{fig:gamma}
    \end{center} 
\end{figure}

One advantage of Bayesian inference is that we can use the posterior results from previous studies to define our prior information. We use the posterior distribution of Paper V to set the prior distributions on the gravitational potential parameters, $\gamma$ and $\Phi_0$. For $\Phi_0$, a gamma distribution was fit with parameters shown in Figure~\ref{fig:phi0}. In Paper V, the gamma distribution on $\gamma$ was narrow and the resulting marginal distribution did not change in light of the GC data. Thus, we used the same prior on $\gamma$ as was used in Paper V (Figure~\ref{fig:gamma}).
\vskip12pt
\begin{table}[htp!]
\begin{center}
\begin{tabular*}{\columnwidth}{@{\extracolsep{\fill}}cll}
\
Model Parameter & Prior & Hyperprior Parameters\\
\tableline
$\Phi_0$ & Gamma & k = 32.4; $\psi$ = 0.69 \\ \tableline
$\gamma$ & Gaussian & $\tau$ = 0.50; $\sigma$ = 0.06 \\ \tableline
$\alpha$ & Gamma & k = 67; $\psi$ = 0.0075 \\ \tableline
$\beta$ & Uniform & a = -3; b = 1 \\ \tableline
\end{tabular*}
\tabcaption{For the Gaussian distributions, the mean is $\tau$, and the standard deviation is $\sigma$. Gamma distributions are described with shape k and rate $\psi$. Uniform distributions have minimum and maximum a and b. \label{table:priors}}
\end{center}
\end{table}

The velocity anisotropy parameter $\beta$ is treated as a free constant variable with uniform prior distribution. The posterior distribution of $\beta$ from Paper V has a lower bound of -1. However, initial tests using the DG data showed this lower bound to be too strict, with the Markov chain wondering into the extreme of -1. Thus, we place a uniform prior with maximum 1 and minimum -3 on $\beta$. The assumption that the velocity anisotropy is constant at all radii and the effect this may have on our study is investigated in Section~\ref{sec:results}.

For the prior on $\alpha$, which describes the spatial distribution of the dwarf galaxies, we relied on simulations to inform our prior distribution.  \cite{Sanderson_20} generated synthetic Gaia DR2 surveys from three MW-like galaxies from the \textit{Latte}  suite (first introduced in \citealt{2016ApJ...827L..23W} of FIRE-2 simulations \citealt{2018MNRAS.480..800H}).   In this study we use MW-like host-halos \textit{i},\textit{f}, and \textit{m} (at z=0). The DM ``subhalos" of the host halos were found using the Rockstar code presented in \citet{Behroozi_13}.  We use subhalos containing stellar mass between $10^5$ and $10^{9} M_{\odot}$ to constrain the prior on $\alpha$.  These limits aim to include the smallest satellite DGs and to exclude any large neighboring bodies; in other words, we want to select subhalos that mimic the DG properties we observe around the MW.  This upper limit of $10^9M_{\odot}$ was also used by \citet[]{GKimmel_19}, who investigated the DGs in the FIRE simulations).  

\begin{figure}[ht!]
    \centering 
    \includegraphics[width=\columnwidth]{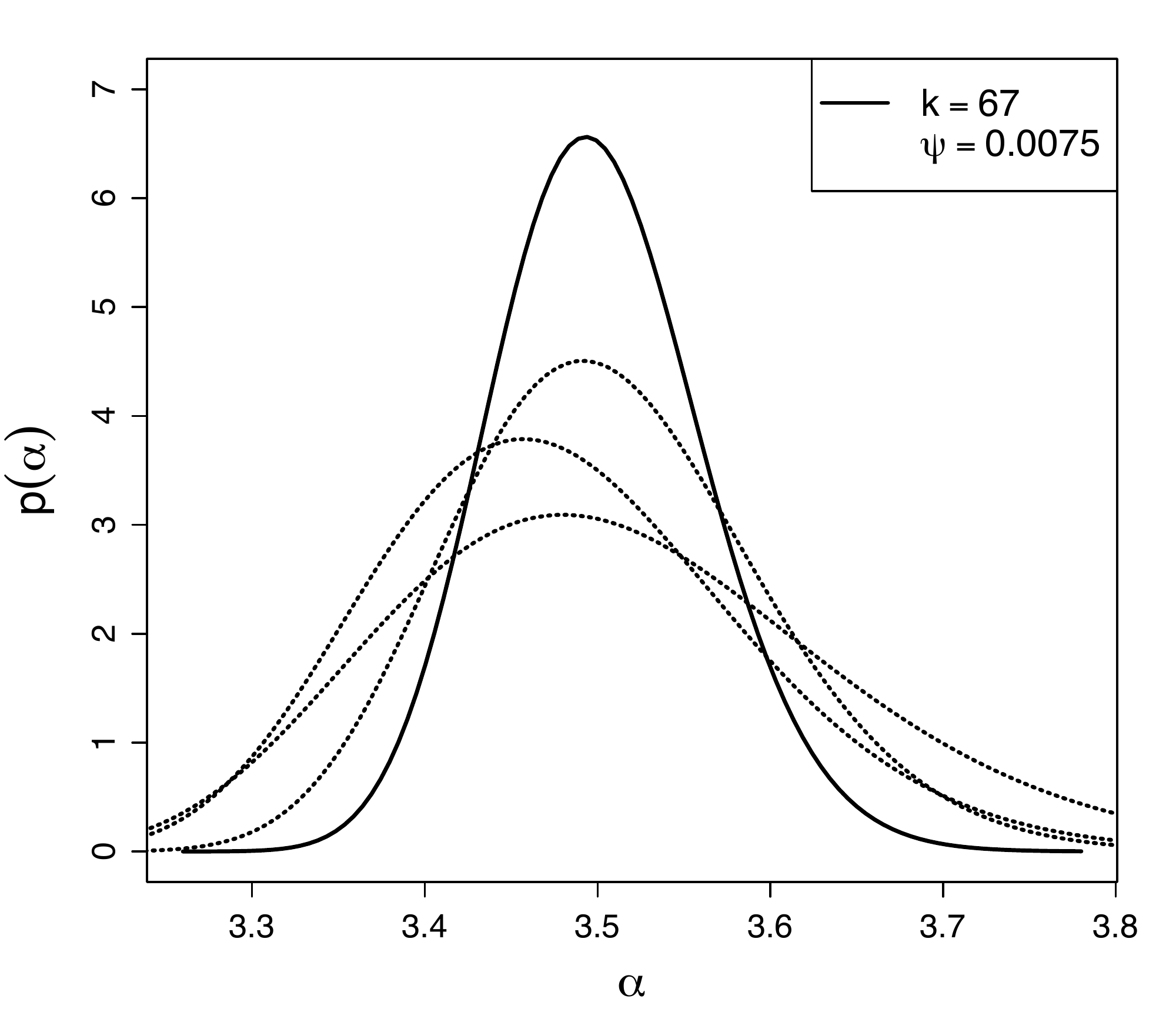}
    \caption{Prior distribution for the tracer power law slope $\alpha$ (solid curve, $N(k, \psi)$), along with the distributions of $\alpha$ from each of the three halos' subhalo populations.}
    \label{fig:alpha}
\end{figure}

For each host halo, we use the total galactocentric distance of each subhalo to find a posterior distribution for $\alpha$, which is a Gamma distribution (the dotted curves in Figure~\ref{fig:alpha}, see \citet{Eadie_16}). Our motivation is to use these posteriors to define a prior on $\alpha$ for this study. Thus, we combine the three subhalo samples to define the final prior on $\alpha$ (the solid curve in Figure~\ref{fig:alpha}). 

\subsection{CMP and Mass Model}\label{sec:massmodel}

With priors placed on the physical model's parameter, and with the DG data, we sample the model and nuissance parameters using the hierarchical Bayesian framework GME. From the Markov chain samples, which are proportional to the posterior distribution, we can calculate mass estimates and produce CMPs. At each Galactocentric distance $r$, we calculate the mass within that $r$ for each set of model parameters $(\Phi_{0,j},\gamma_j)$ from the Markov chain, following
\begin{equation}
\centering
        M(<r) = \frac{ \gamma \Phi_0}{G}\times(\frac{r}{\text{kpc}})^{1-\gamma}.
    \label{eqn:cmp}
    \end{equation}
In this way, we obtain a marginal distribution for $M(<r)$, calculating a CMP with 50, 75, and 95\% Bayesian credible regions (Section~\ref{sec:results}). We also report a virial mass $M_{200}$ and radius $r_{200}$, where $\Delta_c=200$, and assuming a Hubble parameter of $H_0=73\text{ km s}^{-1}\text{Mpc}^{-1}$.

\section{Results \& Discussion}\label{sec:results}
In this section, we first discuss the MW's CMP and $M_{200}$ results from using our full set of DG kinematic data. We then explore the sensitivity of our method's estimates to different subsets of the data. Throughout, results are compared to recent mass estimates from studies using \textit{Gaia} DR2, as well as studies that specifically use DGs.

\subsection{Full Dwarf Galaxy Data Set}\label{sec:fullsetresults}

\begin{figure*}[ht!]
\centering
    \includegraphics[width=1.5\columnwidth]{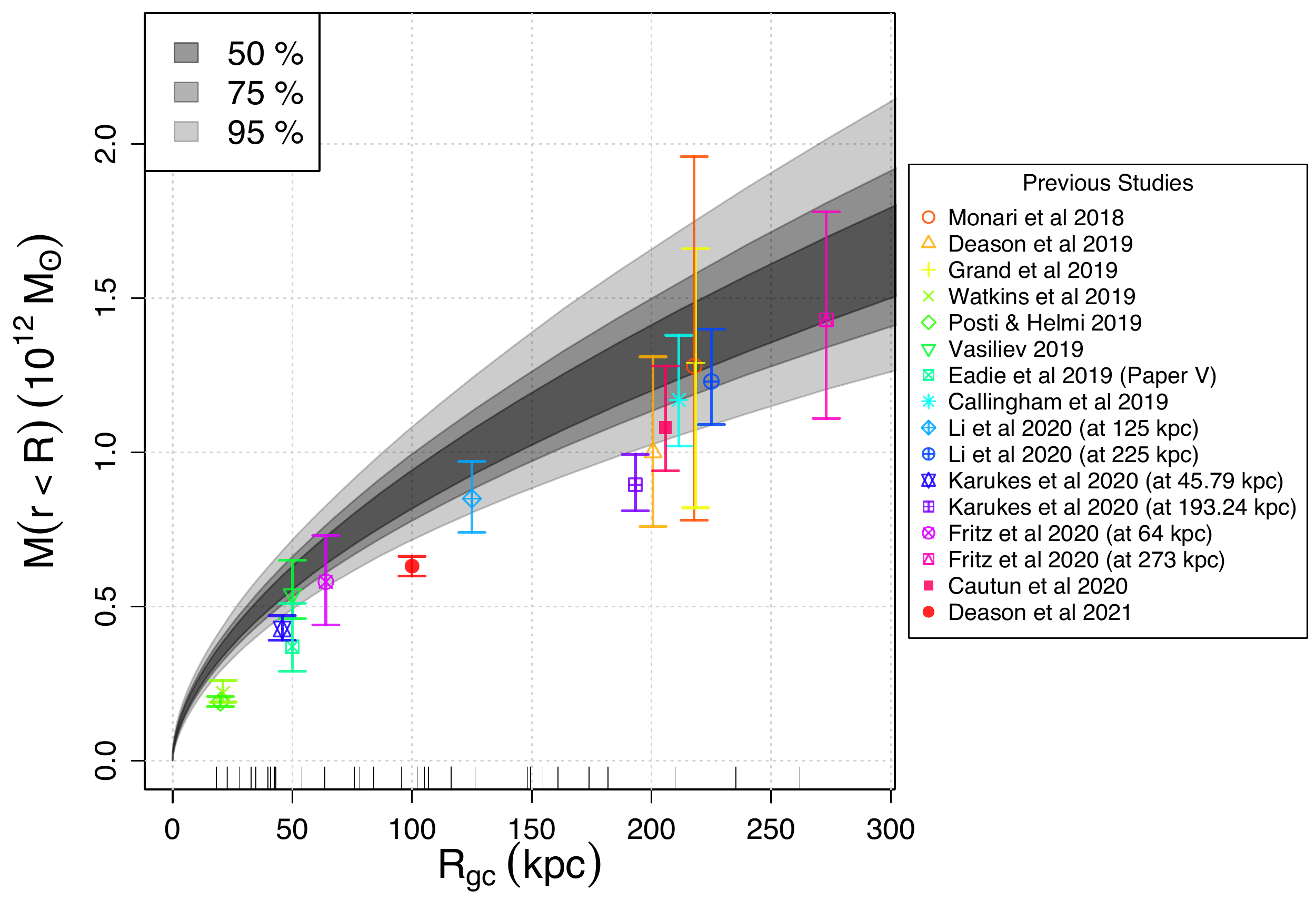}
    \caption{CMP of DG results using the whole dataset, showing locations of DGs, and 50, 75, and 95\% c.r.. Mass estimates with error bars at specific radii from previous studies using \textit{Gaia} DR2 are also plotted and listed.}
    \label{fig:cmp-dg}
\end{figure*}

\begin{figure}[ht!]
    \includegraphics[width=\columnwidth]{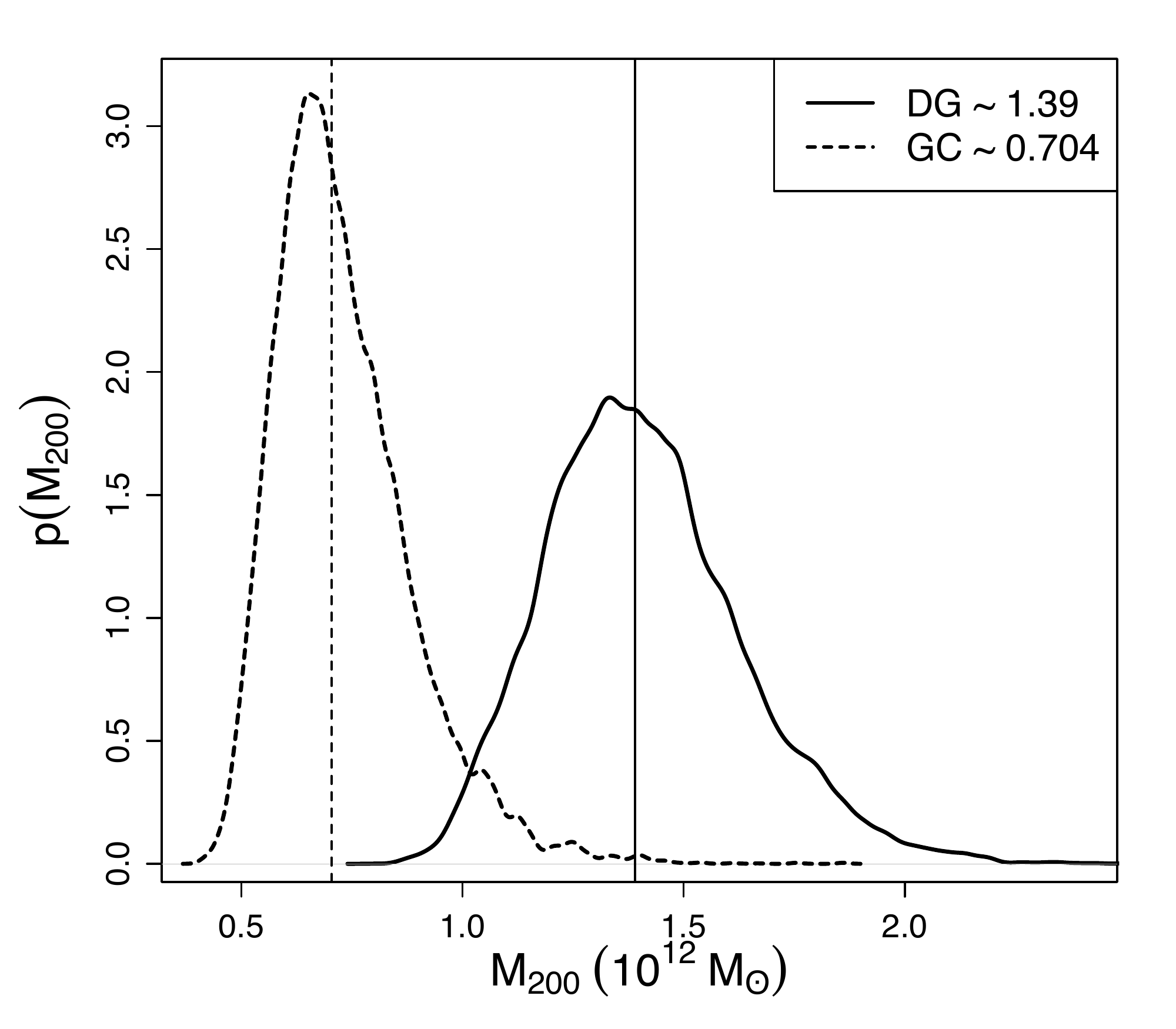}
    \caption{Comparing GC and DG marginal distributions of $M_{200}$ with median values listed in the legend and shown as vertical lines. The mass estimate using DGs is considerably higher than that using GCs in Paper V. The legend shows the approximate median values of $M_{200}$ for the MW mass in solar masses, as estimated by DGs and GCs.}
    \label{fig:dg_gc_margdist}
\end{figure}

\begin{figure*}[hbtp]
\begin{center}
    \includegraphics[width=2\columnwidth]{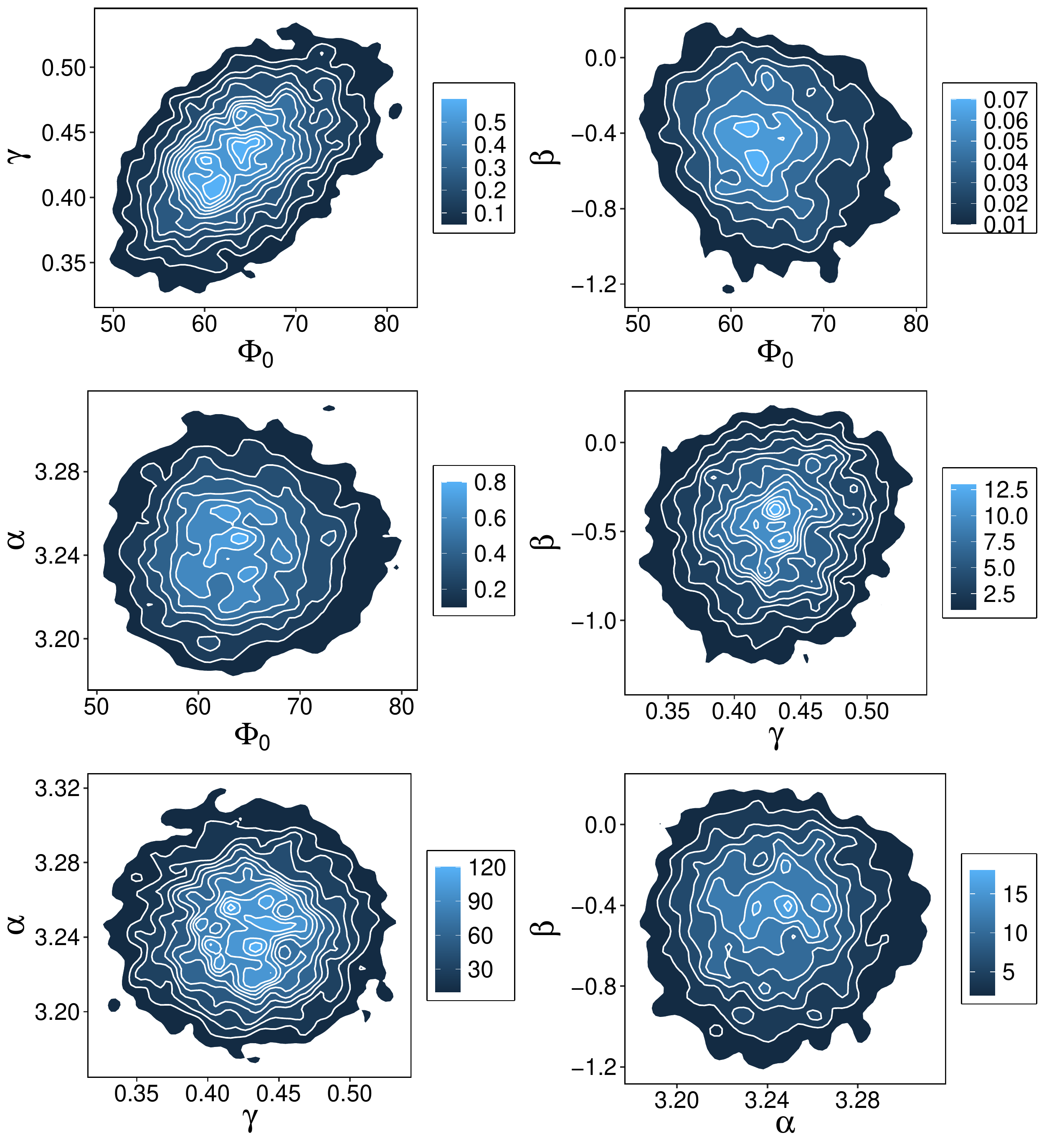}
    \caption{Posterior distributions of the model parameters when using the full set of DG tracers. There is little to no correlation between two of the parameters, except for between $\Phi_0$ and $\gamma$, which show a positive correlation.}
    \label{fig:parplots}
\end{center}
\end{figure*}

Our results for the full set are shown in Figures~\ref{fig:cmp-dg}, \ref{fig:dg_gc_margdist}, \ref{fig:parplots}, and Table~\ref{tab:results}. Fig~\ref{fig:cmp-dg} shows the MW CMP with 50, 75, and 95\% Bayesian c.r. with recent estimates from other literature. For convenience, we also show the mass contained within specific distances in Table~\ref{tab:results} and the marginal posterior distribution of $M_{200}$ in Figure~\ref{fig:dg_gc_margdist}. Note that the estimate within 262 kpc is shown because this is the distance of the farthest DG in our dataset. 

When we compare the estimated CMP to that of Paper V, we find that the 95\% c.r. of the profiles do not overlap at radii beyond $r=30\text{ kpc}$. Specifically, Paper V found median $r_{200}=178.43\text{ kpc}$, $M_{200} = 0.70 (0.51,1.10)\times 10^{12} M_{\odot}$, and $M(r=262\text{ kpc}) = 0.85 (0.63, 1.25)\times 10^{12} M_{\odot}$, which are significantly different than our estimates (Figure~\ref{fig:dg_gc_margdist}). One explanation could be that the GCs are distributed closer within the MW halo than the DGs, and therefore provide a different estimate of $\gamma$.

\begin{figure*}[ht!]
\begin{center}
    \includegraphics[width=\columnwidth]{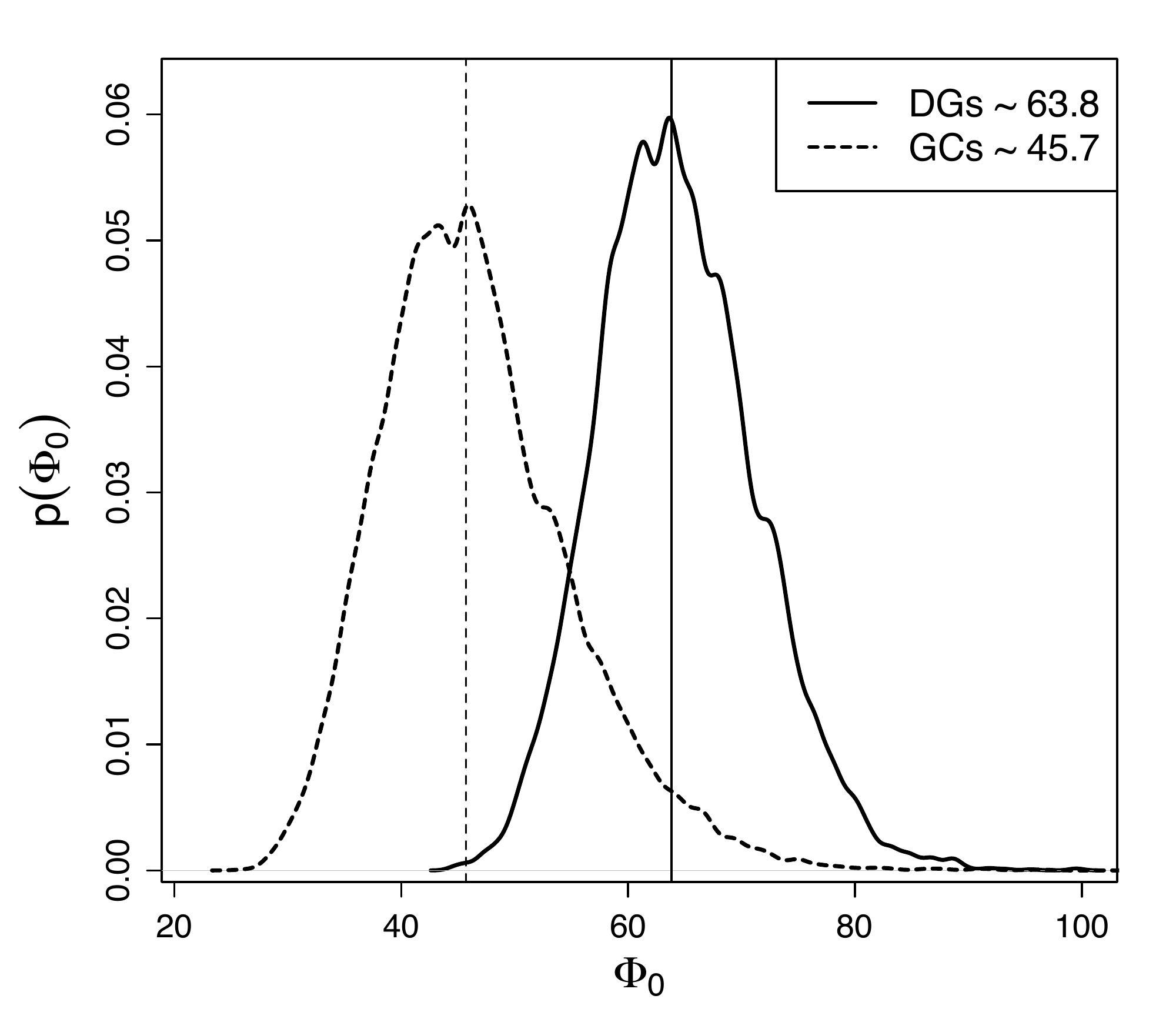}
    \includegraphics[width=\columnwidth]{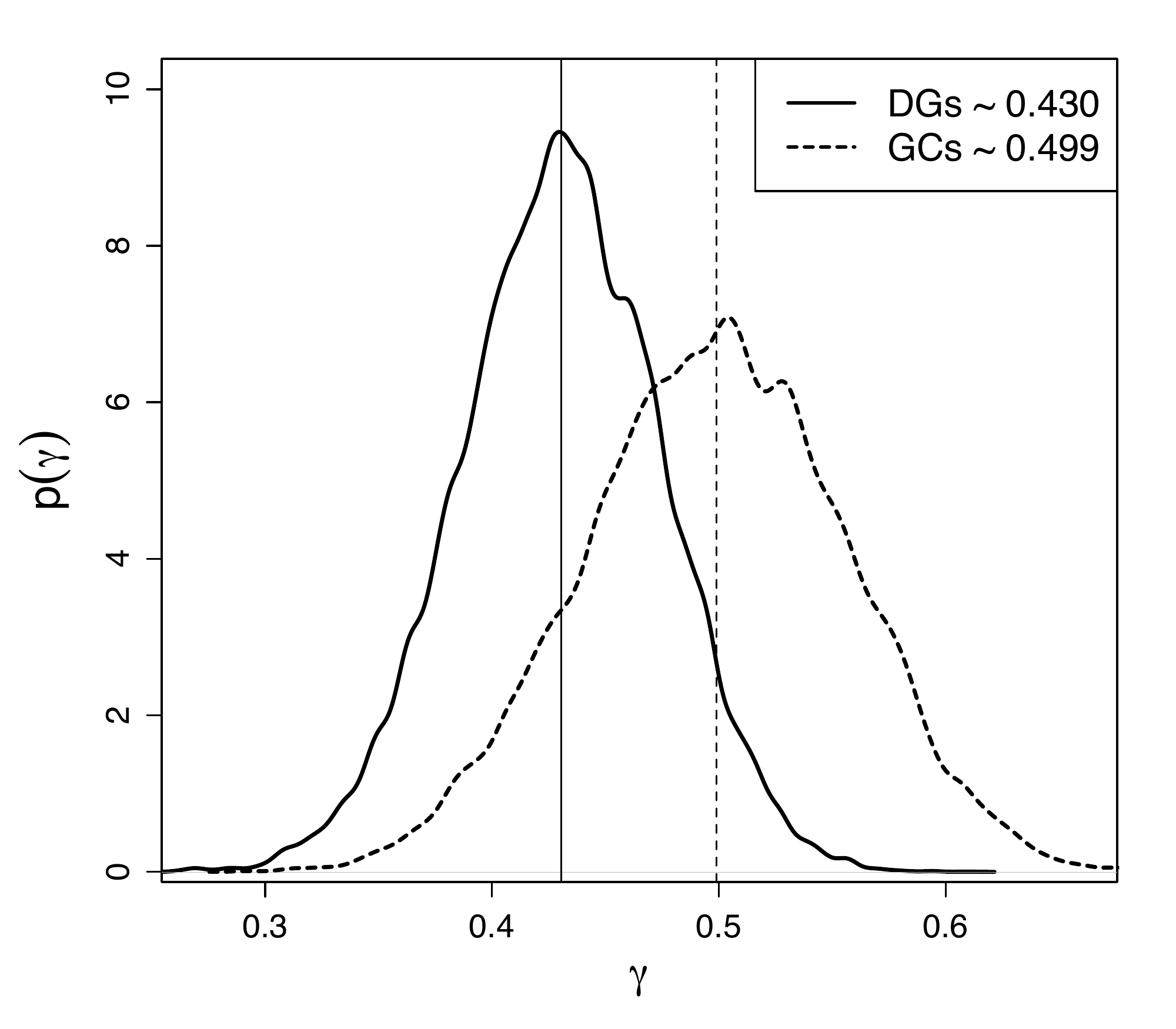}
    \caption{Marginal distributions of $\Phi_0$ and $\gamma$ (solid line) compared to the marginal distribution from Paper V (using GCs, dotted curves). Priors for this study were set using the GC results in Paper V. Thus, through this comparison we observe a significant shift in both parameter estimates of $\Phi_0$ and $\gamma$ given the new DG data.}
    \label{fig:dg_gc_phiandgam}
\end{center}
\end{figure*}

To see how much the new data affects the parameter estimates, we compare the marginal distributions of $\Phi_0$ and $\gamma$ to those from Paper V, and $\alpha$ to its prior distribution. The gravitational potential parameter estimates for the DG results have a larger scale $\Phi_0$ (median 63.8 (100 km/s)$^2$) and smaller slope $\gamma$ (median 0.43) than those in Paper V (Figure~\ref{fig:dg_gc_phiandgam}). Since the priors on these parameters were formed from the results of Paper V (i.e., using GCs), and we were using the same model, the DG tracer data must have caused these parameters to shift in value. Overall, these results suggest that the choice of tracer has a considerable effect on the mass estimate, given the model assumption. 

\begin{figure}[ht!]
    \includegraphics[width=\columnwidth]{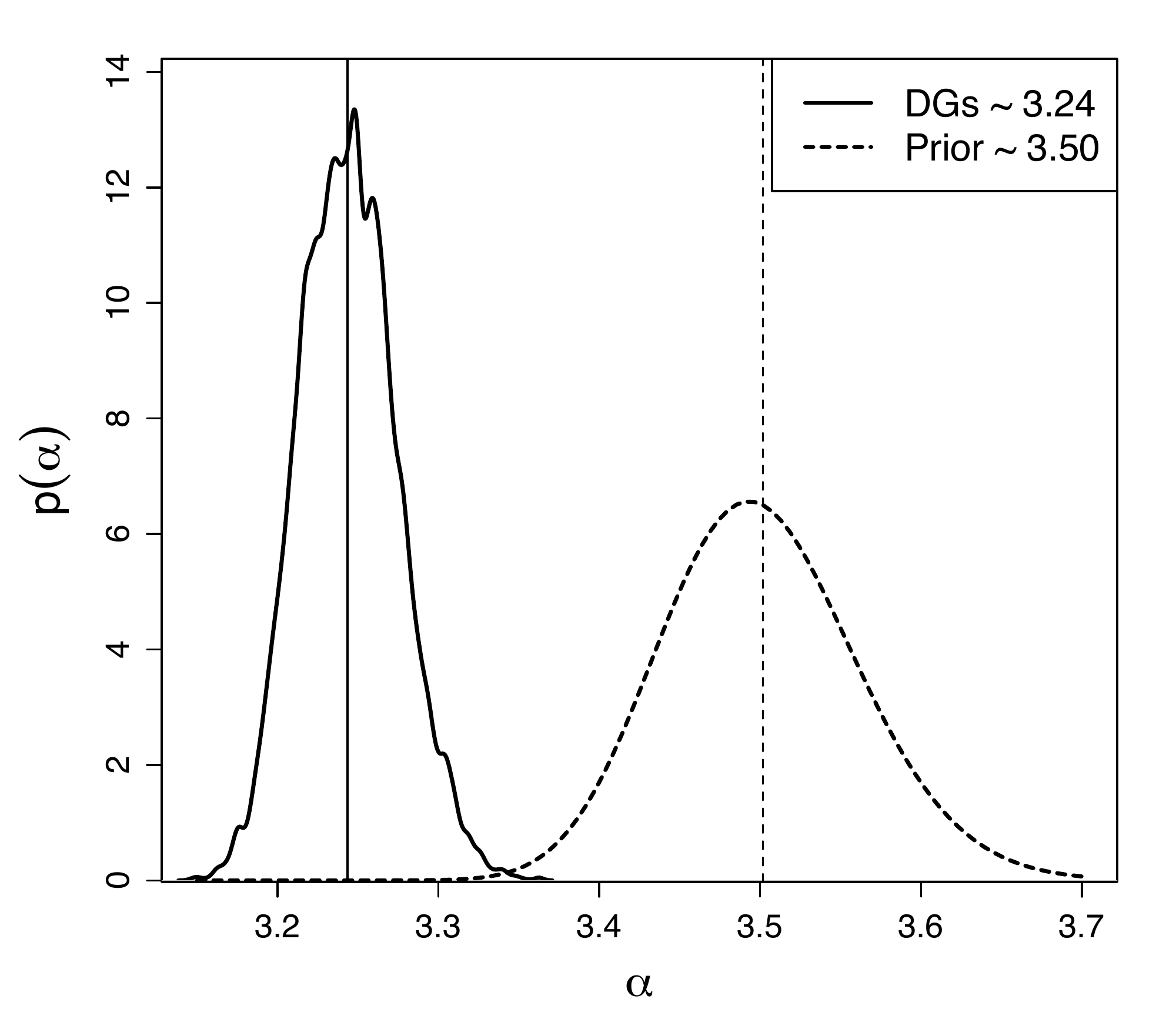}
    \caption{Marginal distribution of $\alpha$ compared to the prior distribution set on $\alpha$. Median values are shown as vertical lines and labelled in the legend.}
    \label{fig:finalalpha}
\end{figure}

The parameter for the spatial distribution of the tracers, $\alpha$, is different for GCs and DGs (Figure~\ref{fig:finalalpha}). As mentioned in Section~\ref{sec:priors}, the prior on $\alpha$ was set using the Ananke simulations of DG-like subhalos. The resulting median $\alpha=3.24$, suggesting a much more shallow distribution of DGs than shown in the Ananke halos. 

\begin{table}[ht!]
\begin{center}
\begin{tabular*}{\columnwidth}{@{\extracolsep{\fill}}lllll}
\toprule
Estimate                & C.r.      & Lower     & Median    & Upper \\ 
\hline
                        & 50\%      & 216.23    &           & 231.71  \\
                        & 75\%      & 211.26    &           & 237.72  \\
$r_{200}$               & 95\%      & 202.70    & 223.87    & 248.41  \\
\hline
                        & 50\%      & 1.25      &           & 1.54  \\
                        & 75\%      & 1.17      &           & 1.66  \\
$M_{200}$               & 95\%      & 1.03      & 1.39      & 1.90  \\
\hline
                        & 50\%      & 0.55      &           & 0.63  \\
                        & 75\%      & 0.53      &           & 0.66 \\ 
$M(r=50\text{ kpc})$    & 95\%      & 0.49      & 0.59      & 0.72  \\ 
\hline
                        & 50\%      & 0.82      &           & 0.94  \\
                        & 75\%      & 0.78      &           & 0.99  \\ 
$M(r=100\text{ kpc})$   & 95\%      & 0.72      & 0.88     & 1.08  \\
\hline
                        & 50\%      & 1.39      &           & 1.66 \\
                        & 75\%      & 1.31      &           & 1.77 \\ 
$M(r=262\text{ kpc})$   & 95\%      & 1.18      &1.52       & 1.96  \\
\hline
\end{tabular*}
\tabcaption{Median estimates and Bayesian c.r. for $r_{200}$, $M_{200}$, and $M(r=50,100,262\text{ kpc})$ found with full dataset. Masses are expressed in $10^{12}M_{\odot}$.\label{tab:results}}
\end{center}
\end{table}

Lastly, we find the median estimate for $\beta$ is $-0.48$, much lower than the GC results. This highlights the different velocity anisotropy of the inner and outer satellite populations. The results of Paper V suggest a \textit{radially} velocity anisotropy of the GC population, whereas the DGs seem to have a more tangential velocity anisotropy. We quickly note that our $\beta$ estimate is similar to \cite{Riley_19}; the estimate of $\beta$, and its sensitivity to the DGs, is discussed further in Section~\ref{sec:subsets}. 

Our median estimate for $M_{200}$ is $1.39(1.03,1.9)\times10^{12}M_{\odot}$ (with 95\% credible interval). This is larger than that found with GCs in Paper V, but is still comparable to recent results.

The full set's estimate is similar to \citet[]{Monari_18} and \citet{Hattori_18}, each finding $M_{200}>1.4\times10^{12}M_{\odot}$. However, \citet[]{Fritz_2020} found 68.3\% c.r. $M(<273\text{ kpc})=1.43^{+0.35}_{-0.32}\times10^{12}M_{\odot}$, which is slightly lower than our extrapolated result of $M(<273\text{ kpc})=1.56(1.37,1.77)\times10^{12}M_{\odot}$
, but very similar to the result found in the \textgreater45 kpc subset, $M(<273\text{ kpc})=1.37(1.19,1.59)\times10^{12}M_{\odot}$,
discussed in Section~\ref{sec:subsets}.
\citet{Li_2020}, also using the Riley catalog, applied a simulation-based distribution function method to estimate 68\% c.r. $M_{200}=1.23^{+0.21}_{-0.18}\times10^{12}M_{\odot}$. Their method did not assume all satellites are bound to the MW, and their estimates were found to be unbiased from the simulations and unaffected by the LMC. Their estimate is similar to the results of the \textgreater45 kpc subset (described next) of $M_{200}=1.19(1.00,1.42)\times10^{12}M_{\odot}$. 

\subsection{Sensitivity Analyses}\label{sec:subsets}

While setting up our analysis, we removed four LMC-associated satellites from the DG data set: CarinaII, Carina III, Horologium I and Hydrus I (Section~\ref{sec:data}). To justify removing these DGs from our analysis, we ran a test including these four satellites. The median $M_{200}$ estimate increased to $1.60\times10^{12}M_{\odot}$. Their influence on the mass is not surprising; our model assumes that DGs are bound to the MW, and any DGs that are unbound (and/or associated with the LMC) can lead to a larger mass estimate. This was also observed in simulation tests of Paper IV. 

At the same time, it has been suggested that a \textit{massive} LMC could affect other DGs too. In this section, we explore this possibility and re-run our analysis on various subsets of DGs defined by distance cut-offs, spatial octant choices, and a jackknife analysis. Our different subset choices are motivated by previous studies, as described below. 

\cite{Erkal_20} argue that the LMC pulls the MW out of equilibrium, leading to bulk motion in outer tracers (beyond 30 kpc) compared to inner tracers, and suggest that ignoring the LMC's effect can lead to an overestimate in MW mass up to 50\%. \textit{Thus, one subset of DGs to consider analysing are those located beyond 30 kpc.} 

\cite{Garavito_Camargo_19} showed that interactions between the MW and an infalling, massive LMC could created a wake of overdensities and distinct kinematic patterns beyond 45 kpc. \textit{Thus, a second subset of DGs to consider are those located beyond 45 kpc.}

Based on the evidence found in the aforementioned studies, we ran our analysis on five subsets according to Galactocentric distance $R$: ($>$30kpc, $>45$kpc, $<45$kpc, and between 30 and 45 kpc). Note that the subset of DGs within 30 kpc only contains four DGs. Estimating four parameters with four data points would not provide much insight, so we do not analyse this subset. 

\begin{figure}
\begin{center}
    \includegraphics[width=\columnwidth]{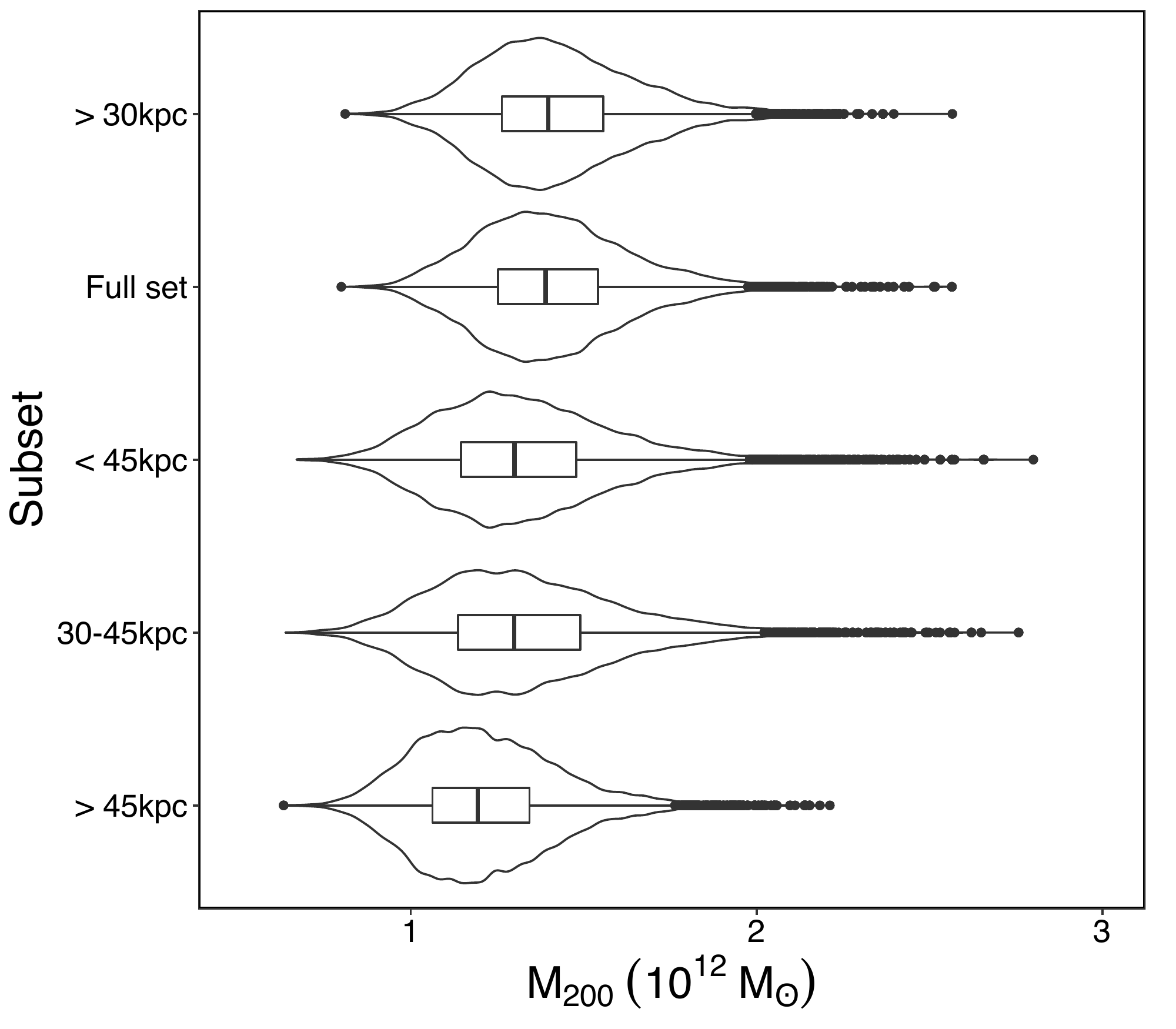}
    \caption{$M_{200}$ violin plots for each subset (see Table \ref{tab:subsets}). The distributions are shown in decreasing order from top to bottom; the subset that suggests the largest $M_{200}$ is $>30$kpc, while the subset that suggests the lowest $M_{200}$ is $>45$kpc.}
    \label{fig:M200 comp}
\end{center}   
\end{figure}

The five subsets and the median mass estimate of the MW $M_{200}$ for each are shown in Table~\ref{tab:subsets}, with columns in order from highest to lowest median. Their marginal distributions are shown in Figure~\ref{fig:M200 comp} as violin plots.

\citet{Erkal_20} also found that using satellites from certain spatial octants around the MW led to biased mass estimates when there was a large, massive LMC present ($1.5\times10^{11}M_{\odot}$). The octants are defined by breaking the Galaxy into the positive and negative regions along the X, Y and, Z axes. In \citet{Erkal_20}, the largely unbiased octants are, in (X,Y,Z): (-,-,+), (+,+,-), (+,-,+), and (-,+,+), which we call Group 1, whereas the biased octants are: (-,-,-), (-,+,-), (+,-,-), and (+,+,+), which we call Group 2. In our full data set, there are 21 DGs in Group 1 and 11 DGs in Group 2. After applying the method to these two subsets, Group 1 produced a mass estimate of $M_{200}=1.32 (0.96, 1.84)\times10^{12}M_{\odot}$ and Group 2 produced a mass estimate of $M_{200}=1.19 (0.83, 1.76)\times10^{12}M_{\odot}$ (95\% c.i.). Although the mass estimates do differ somewhat, the 95\% credible regions overlap significantly. Additionally, we note that none of the subsets listed in Table~\ref{tab:subsets} are dominated by either Group 1 or Group 2 DGs. 

We also perform a jackknife analysis on other DGs that are suggested to be either interacting with the LMC or gravitationally unbound to the MW.  \cite{Patel_20} showed that Aquarius II and Sculptor may have interacted with the LMC, and \cite{Bajkova_20} found that Aquarius II, Grus I, Hydra II, Leo IV, Leo V and, Pisces II may not be gravitationally bound to the MW. Additionally two DGs Carina and Fornax are suggested to be LMC associated \citep{Jahn_2019,Pardy_19}. Each DG was removed one by one in a jackknife analysis from the \textgreater45 kpc subset, and we found that no single DG made any significant impact on the resulting mass estimate.

\begin{table*}[ht!]
\begin{center}
\begin{tabular}{lccccccc}
\toprule
Dwarf Galaxy & {R\textgreater30 kpc} & {Full Set} & {R \textless45 kpc} & {30 kpc\textless\ R \textless 45 kpc} & {R \textgreater 45 kpc} & \\ \hline
Aquarius II & X & X & &  & X & B+P\\ \hline
Bootes I & X & X & & & X & \\ \hline
Bootes II & X & X & X & X & & B\\ \hline
Canes Venatici I & X & X & & & X & \\ \hline
Canes Venatici II & X & X & & & X & \\ \hline
Carina I & X & X & & & X &\\ \hline
Coma Berenices I & X & X & X & X & &\\ \hline
Crater II & X & X & & & X & \\ \hline
Draco I & X & X & & & X &\\ \hline
Draco II & & X & X & & & \\ \hline
Fornax & X & X & & & X & \\ \hline
Grus I & X & X &  &  & X& B\\ \hline
Hercules & X & X &  & & X& \\ \hline
Hydra II & X & X &  &  & X&B\\ \hline
Leo I & X & X & & & X &\\ \hline
Leo II & X & X & & & X &\\ \hline
Leo IV & X & X &  &  & X& B\\ \hline
Leo V & X & X &  &  & X &B\\ \hline
Pisces II & X & X &  &  & X & B\\ \hline
Reticulum II & X & X & X & X && P\\ \hline
Sagittarius I & & X & X & && \\ \hline
Sculptor & X & X &  &  & X& P\\ \hline
Segue 1 &  & X & X &  & &P\\ \hline
Segue 2 & X & X & X & X & &\\ \hline
Sextans & X & X & & & X &\\ \hline
Triangulum II & X & X & X & X & &\\ \hline
Tucana II & X & X & & & X & \\ \hline
Tucana III &  & X & X &  & &P\\ \hline
Ursa Major I & X & X & & & X&\\ \hline
Ursa Major II & X & X & X & X & &  \\ \hline
Ursa Minor & X & X & &  & X &\\ \hline
Willman 1 & X & X & X & X & &\\ \hline\hline
Total DGs  & 28 & 32 & 11 & 7 & 21 & \\ \hline
Median $M_{200}$ & 1.40 & 1.39 & 1.30 & 1.30 & 1.19 \\ \hline
50\% c.r. & (1.26, 1.56) & (1.25, 1.54) & (1.15, 1.48) & (1.14, 1.49) & (1.06, 1.34) \\ \hline
95\% c.r. & (1.04, 1.93) & (1.03, 1.90) & (0.90, 1.93) & (0.88, 1.94) & (0.87, 1.68) \\ \hline
\end{tabular}
\caption{The subsets tested with their included DGs. The rightmost column shows DGs which \cite{Bajkova_20} (B) found to be not gravitationally bound to the MW, and which DGs \citet{Patel_20} (P) suggested may have interacted with the LMC (Aquarius II is mentioned in both). At the bottom of the table, the computed $M_{200}$ median values are included with 50\% and 95\% credible regions, each in units of $10^{12} M_{\odot}$.\label{tab:subsets}}
\end{center}
\end{table*}

We return now to discuss the results listed in Table~\ref{tab:subsets} and their $M_{200}$ distributions (Figure~\ref{fig:M200 comp}). In short, we find that the DGs between 30 and 45kpc seem to have the most influence on the mass estimate. We expected to find a difference in mass estimates with distant DGs versus close DGs, potentially related to either disequilibrium as discussed in \cite{Erkal_20} or the LMC infall wake as discussed in \cite{Garavito_Camargo_19}. However, the \textless45 kpc and \textgreater45 kpc subsets produced rather similar results (a $0.11\times10^{12}M_{\odot}$ difference). Instead, the most significant difference is between the \textgreater45 kpc and the \textgreater30 kpc subset (a difference of $0.21\times10^{12}M_{\odot}$), as observed in the top and bottom violin plots in Figure~\ref{fig:M200 comp}. We also note that the four DGs within 30 kpc have little impact on the mass estimate --- the full set and the \textgreater30 kpc give almost identical $M_{200}$ estimates. Moreover, the \textless45 and $30-45$kpc subsets have virtually identical estimates. This result, along with our other sensitivity tests, implies that DGs between 30 kpc and 45 kpc may have some influence on MW mass estimates.

To further investigate the influence of DGs located between 30 and 45 kpc, we looked at DGs mentioned by \citet{Patel_20} and \citet{Bajkova_20} that are within this range --- i.e., Reticulum II and Bootes II. After removing Reticulum II, which is thought to be recently captured by the LMC, there is virtually no difference in the mass estimate. Removing Bootes II, which was suggested to be unbound, does produce a slightly different mass estimate (1.30 with and 1.22 without), but there is significant overlap in the 50$\%$ credible region.

\begin{figure}
\begin{center}
    \includegraphics[width=\columnwidth]{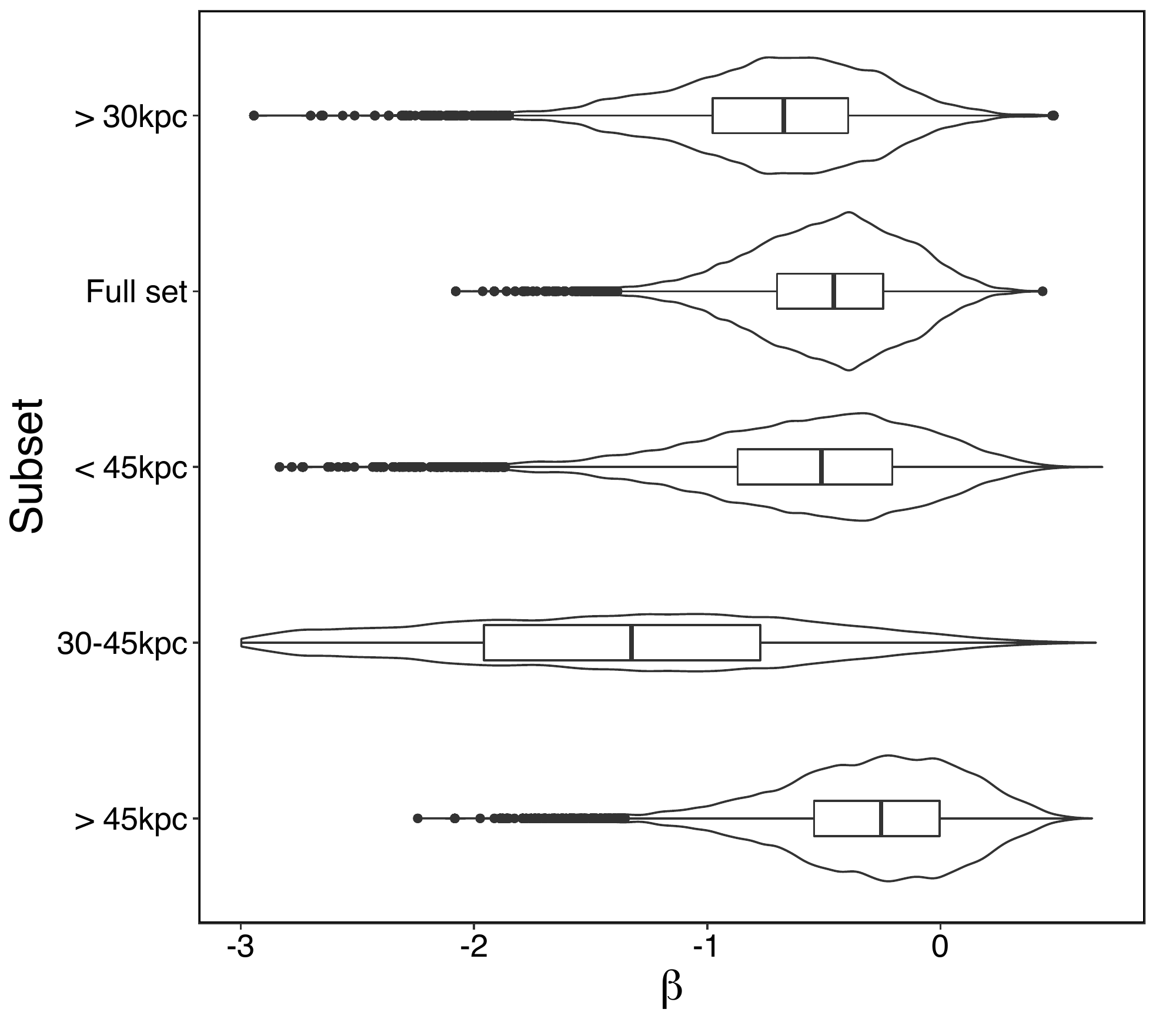}
    \caption{Violin plots for $\beta$, shown in the same order as Figure~\ref{fig:M200 comp}. The standard deviations for $\beta$, from top to bottom are 0.33, 0.40, 0.43, 0.78, and 0.49.}
    \label{fig:Beta comp}
\end{center}
\end{figure}

Figure~\ref{fig:Beta comp} shows the velocity anisotropy marginal distributions for the various subsets. Clearly, $\beta$ is poorly constrained for the 30-45 kpc subset. It is possible that the small number of DGs (seven) within this subset are causing the uncertainty. However, the \textless45 kpc subset, which has only eleven DGs, is not as uncertain. Moreover, when we initially looked at the \textless30 kpc subset (with just four DGs), the uncertainty in $\beta$ was similar to the other subsets and reasonably well-estimated --- therefore, the small number of DGs between 30-45kpc may not account for the large uncertainty of $\beta$ in this region. Apart from the 30-45 kpc subset, similarly to \cite{Riley_19}, our results show more negative $\beta$ values for the inner MW regions and more positive $\beta$ values for outer MW regions.

Overall, the findings in Figure~\ref{fig:Beta comp} may shed some light on the significance of the seven DGs within 30-45 kpc. Is it possible that the LMC is massive, and caused a disruption in motion in this region? Further exploration in the region between 30 and 45kpc is needed, and incorporating a mixed data set of GCs, DGs, and halo stars together to increase the sample size in this region might yield some insight. We leave this to future work.

\begin{figure*}
\begin{center}
    \includegraphics[width=1.5\columnwidth]{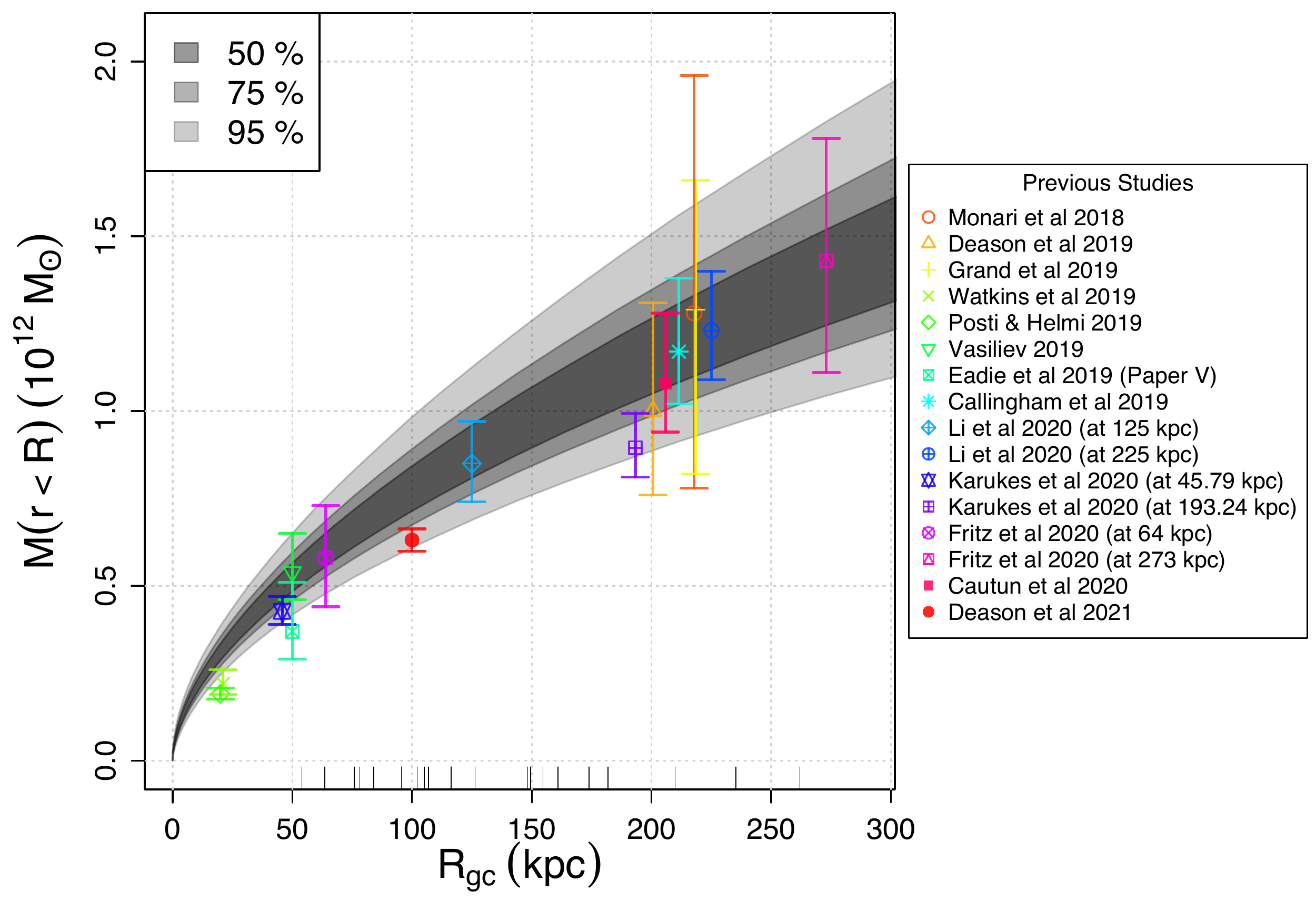}
    \caption{Results for the CMP, showing 50, 75, and 95\% Bayesian c.r, when only using DGs beyond 45 kpc. Other recent studies using \textit{Gaia} DR2 have masses plotted at certain radii for comparison.}
    \label{fig:finalcmp}
\end{center}
\end{figure*}

Keeping sensitivity analysis results in mind, we report the mass for the full set of DGs \textit{and} for the subset beyond 45 kpc as our final results. The median model parameters of the latter are $\Phi_0=56.48$, $\gamma=0.43$, $\alpha=3.28$, and $\beta=-0.26$, while $M_{200}=1.19(0.87,1.68)$ (95\% c.i.). This shows a decreased $\Phi_0$, a less tangential $\beta$, and a lower mass than when using the full DG data set. We note these results are very similar to others using \textit{Gaia} DR2's DGs (\citet{Li_2020,Fritz_2020}), and the 95\% overlaps greatly with the CMP of Paper V at all radii.  For future work and reference to our mass estimates we recommend referencing and using the values in Figure \ref{fig:finalcmp} and Table \ref{tab:finalresults}.

\begin{table}
\begin{center}
\begin{tabular*}{\columnwidth}{@{\extracolsep{\fill}}lllll}
\toprule
Estimate                & C.r.      & Lower     & Median    & Upper \\ 
\hline
                        & 50\%      & 204.71    &           & 221.33  \\
                        & 75\%      & 199.66    &           & 227.25  \\
$r_{200}$               & 95\%      & 191.12    & 212.80    & 238.44 \\
\hline
                        & 50\%      & 1.06      &           & 1.34  \\
                        & 75\%      & 0.99      &           & 1.45  \\
$M_{200}$               & 95\%      & 0.87      & 1.19      & 1.68  \\
\hline
                        & 50\%      & 0.48      &           & 0.57  \\
                        & 75\%      & 0.46      &           & 0.60 \\ 
$M(r=50\text{ kpc})$    & 95\%      & 0.42      & 0.52      & 0.65  \\ 
\hline
                        & 50\%      & 0.71      &           & 0.84  \\
                        & 75\%      & 0.67      &           & 0.89  \\ 
$M(r=100\text{ kpc})$   & 95\%      & 0.61      & 0.78      & 0.98  \\
\hline
                        & 50\%      & 1.22      &           & 1.48 \\
                        & 75\%      & 1.14      &           & 1.58 \\ 
$M(r=262\text{ kpc})$   & 95\%      & 1.02      & 1.34      & 1.78  \\
\hline
\end{tabular*}
\tabcaption{Median estimates and Bayesian c.r. for $r_{200}$, $M_{200}$, and $M(r<50,100,262\text{ kpc})$, found when using only DGs beyond 45 kpc. Masses are expressed in $10^{12}M_{\odot}$.\label{tab:finalresults}}
\end{center}
\end{table}

\section{Conclusions \& Future Work}\label{sec:conc}

In this work, we set out to answer three questions listed in the introduction. To answer these questions, we applied the hierarchical Bayesian method GME to new proper motion data of DGs from Gaia DR2, while also incorporating prior information from both previous studies and simulations. A summary of our answers are below:

\begin{enumerate}
    \item \textit{What is the MW cumulative mass profile (CMP) when using DG data?} \\ The CMP estimates and Bayesian credible regions are shown as the grey bands in Figures~\ref{fig:cmp-dg} (for the entire data set) and \ref{fig:finalcmp} (for DGs beyond 45kpc). The $M_{200}$ estimates in units of $10^{12}M_{\odot}$ for these two cases, and their 95\% credible regions, are 
    \begin{eqnarray*}
        M_{200, \text{full set}} &=& 1.39~(1.03,~1.90) \times 10^{12} M_{\odot} \\
        M_{200, >\text{45kpc}} &=& 1.19~(0.87,~1.68) \times 10^{12} M_{\odot}\\
    \end{eqnarray*}
    Thus, excluding DGs within 45kpc decreased our mass estimate by about 14\%. However, our sensitivity analysis showed that this was caused mainly by DGs between 30 and 45kpc.
    
    \item \textit{Does our model still produce lower estimates when using DG data?} \\
    No, we find that the total mass estimate when using DGs is significantly higher than when GCs are used as tracers (as in Paper V). This is interesting, given that the prior information on model parameters $\Phi_0$ and $\gamma$ were set using the results from the GC study by \cite{Eadie_2019}, which also use the GME code. The model parameter estimates for the full set and the subset of DGs beyond 45kpc are shown in Tables~\ref{tab:results} and Table~\ref{tab:finalresults}.
    
    \item \textit{How sensitive are the mass estimates to particular DGs that could be affected by a massive LMC?} \\
    Our sensitivity analysis (Section~\ref{sec:subsets}) was performed by separating the DG data into subsets determined by distance, by octant, and also by a jacknife analysis. The summary of our sensitivity analysis is shown in Table~\ref{tab:subsets} and Figures~\ref{fig:M200 comp} and \ref{fig:Beta comp}. Overall, we find that our hierarchical Bayesian framework for estimating the mass of the MW is sensitive to tracer choice (i.e., GCs or DGs). We also find that the MW mass estimate depends on which DGs are used, and is related to their distance and location. In particular, there is some uncertainty in the mass due to the DGs between 30 and 45kpc --- including these DGs increased the mass estimate of the MW. If these DGs were indeed affected by a massive LMC then they are increasing the mass estimate of the MW as predicted by \cite{Erkal_20}.
    For the latter effect, however, it is unclear whether these properties are the result of a massive LMC interacting with the DGs or simply randomness due to small sample size.

\end{enumerate}

Next, we wish to reiterate our findings related to the 30 to 45 kpc subset of DGs. The results in Section~\ref{sec:subsets} suggest that DGs between 30 and 45 kpc have an impact on our Galaxy mass estimate. Figure~\ref{fig:Beta comp} in particular shows the inability to estimate $\beta$ for DGs within this region compared to the other subsets. However, due to limited sample size in the DG subsets, we are tentative in making any strong conclusions about whether these differences are the result of a massive LMC. Nevertheless, given the effect of DGs between 30 and 45kpc, we recommend further study of the population of tracers in this region. A study which combines data from halo stars, GCs, and DGs in this region to estimate the mass, and compares this to mass estimates from tracers in other regions, might provide more insight.  

There are many avenues of future work. For example, one could apply this methodology directly to the simulated data from \texttt{latte} to test the ability of the method to recover the true total mass profile of the MW-like host galaxies. A similar such study has already been completed \citep{Eadie_18} with MUGS2 MW-like galaxies, but further testing of the methodology with a different set of simulations is welcome. We leave this to future work. Significant improvements to the GME framework are also underway, which will improve computational  efficiency and which will be applied to data from MW halo stars (Shen et al, in prep).

Another avenue of investigation is to allow the center of the MW to vary with respect to the central potential. By treating this center-offset as a parameter in the model, one could infer the amount of offset implied by the tracer population. 

Now that \textit{Gaia} EDR3 has been released, updated and more precise proper motions for these DGs and other tracer objects objects could be used in a similar analysis. Results from all of these studies can be compared easily if they are reported with CMPs, with the caveat that extrapolations beyond the extent of the data should be taken with care.

This work has made use of data from the European Space Agency (ESA) mission {\it Gaia} (\url{https://www.cosmos.esa.int/gaia}), processed by the {\it Gaia} Data Processing and Analysis Consortium (DPAC, \url{https://www.cosmos.esa.int/web/gaia/dpac/consortium}). Funding for the DPAC has been provided by national institutions, in particular the institutions participating in the {\it Gaia} Multilateral Agreement. This research has made use of NASA's Astrophysics Data System Bibliographic Services.

\acknowledgements
GE acknowledges funding from a NSERC Discovery Grant and Connaught New Researcher Award Grant that helped fund this research.
AW received support from NSF CAREER grant 2045928; NASA ATP grants 80NSSC18K1097 and 80NSSC20K0513; HST grants GO-14734, AR-15057, AR-15809, GO-15902 from STScI; a Scialog Award from the Heising-Simons Foundation; and a Hellman Fellowship.

\software{R Statistical Software Environment \citep{RSoftware} and the following R packages:  \texttt{Cairo} \citep{cairo}, \texttt{ggplot2} \citep{wickamggplot2},\texttt{xtable} \citep{xtable}}

\bibliographystyle{aasjournal}
\bibliography{maindraft}

\end{document}